To Appear in: Itiel E. Dror & Stevan Harnad (Eds) *Cognition Distributed: How Cognitive Technology Extends Our Minds*. Amsterdam: John Benjamins.

---

# Offloading Cognition onto Cognitive Technology

**Itiel E. Dror**
School of Psychology
University of Southampton
Southampton , United Kingdom SO171BJ

**Stevan Harnad**
Chaire de recherché de Canada
Institut des sciences cognitives
Université du Québec à Montréal
Montréal, Québec, Canada H3C3P8
&
School of Electronics and Computer Science
University of Southampton
Southampton , United Kingdom SO171BJ

**ABSTRACT:** "Cognizing" (e.g., thinking, understanding, and knowing) is a *mental state*. Systems without mental states, such as cognitive technology, can sometimes contribute to human cognition, but that does not make them *cognizers*. Cognizers can offload some of their cognitive functions onto cognitive technology, thereby extending their performance capacity beyond the limits of their own brain power. Language itself is a form of cognitive technology that allows cognizers to offload some of their cognitive functions onto the brains of other cognizers. Language also extends cognizers' individual and joint performance powers, distributing the load through interactive and collaborative cognition. Reading, writing, print, telecommunications and computing further extend cognizers' capacities. And now the web, with its network of cognizers, digital databases and software agents, all accessible anytime, anywhere, has become our "Cognitive Commons," in which distributed cognizers and cognitive technology can interoperate globally with a speed, scope and degree of interactivity inconceivable through local individual cognition alone. And as with language, the cognitive tool par excellence, such technological changes are not merely instrumental and quantitative: they can have profound effects on how we think and encode information, on how we communicate with one another, on our mental states, and on our very nature.

**Introductory Overview:**

With the development and wide use of cognitive technologies (Dror, 2007; Dascal & Dror, 2005), questions arise as to their effects on their human users and society, as well as on their own scope and limits: Can cognitive technologies (i) increase cognitive capacities and thus enhance human efficiency? (ii) affect how people and society go about achieving their goals? (iii) highlight and transform how we view ourselves and our goals? (iv) modify how we cognize and thus change our mental states and nature? (v) give rise to new forms of cognition (such as distributed cognition) and mental states that are either distributed across or even embodied in cognitive technology?

These issues are examined as follows:

(1) The notion of an "extended mind" -- with mental states (i.e., felt states) “distributed” beyond the narrow bounds of the individual brain – is not only as improbable as the notion that the US government can have a distributed migraine headache, but arbitrary.

(2) "Cognition" -- if it is simply defined as the ability to do the kinds of things that cognizers like us can do, plus the underlying functional mechanisms for doing them -- can be arbitrarily defined to be as wide or as narrow as we like.

(3) Vagueness about the nature, locus and scope of cognizing leads to a dissociation of “cognitive states” from mental states. However, their co-occurrence had been our only basis for distinguishing cognitive performance capacity from other capacities and functionality (animate or inanimate, narrow or wide).

(4) If cognitive states are indeed not mental states, it follows that "cognitive technology" is not just something *used* by cognizers, but a functional part of the cognitive states themselves, because the boundary between user and tool disappears, and cognitive states become merely instances of functional states in general.

(5) We then do not need the terms "cognitive" and "distributed cognition" at all, and can just talk about relatively complex and wide or narrow functional states, leaving it a coincidence and mystery (at least at this stage) that every single case of what we used to call “cognitive” also happened to be mental.

(6) A way to resolve this is to accept that only mental states are cognitive states, that cognition is only narrow, and that the only place it is "distributed" is within a single cognizer's brain.

(7) The only kind of "technology" that might really turn out to be *intrinsically cognitive*, rather than just being a tool used by cognizers, would be a robot that could pass the Turing Test (TT) -- because such a TT-scale robot would almost certainly have mental states, and hence it would be a cognizer in its own right.

(8) Whatever distributed activity was going on within the functional mechanism generating such a TT robot's performance capacity would then indeed be a case of distributed cognition (exactly as the distributed activity within our own brains is distributed cognition) – even if not all the components of its generating mechanism were located inside the robot's head.

(9) The "cognitive technology" used by such a TT robot, however, would still not be part of its distributed cognitive (hence mental) state, just as it is not a part of ours.

(10) Nor would a group of such TT robots, interacting and collaborating, be a case of distributed cognition; it would merely be a case of collaborative cognition among individual (narrow) TT-robot cognizers, just as it is in the case of a group of collaborating human cognizers.

(11) Cognitive technology does, however, extend the scope and power of cognition, exactly as sensory and motor technology extends the scope and power of the bodily senses and movement.

(12) Just as we can *see* further with telescopes, move faster with cars, and do more with laser microsurgery than we can do with just our unaided hands and heads, so we can *think* faster and further, and do more, with language, books, calculators, computers, the web, algorithms, software agents, plus whatever is in the heads of other cognizers.

(13) Both sensorimotor technology and cognitive technology extend our bodies' and brains' performance capacities as well as giving us the *feeling* of being able to do more than just our bodies and brains alone can do.

(14) Sensorimotor and cognitive technology can thus generate a perceptual change, rather like virtual reality (VR), making us feel a difference in our body image and causal power (perhaps not unlike what the physical metamorphosis from caterpillar to butterfly might feel like, as one sensed one's newfound somatic capacity to fly).

(15) This change in perceived body image is indeed a change in mental state; but although its distal inputs and outputs certainly extend wider than the body (as all sensory inputs and all motor outputs do), the functional mechanism of that altered mental state is still just proximal -- skin and in – exactly as when it is induced by VR technology.

(16) Hence, although sensorimotor and cognitive technology can undeniably extend our bodies' sensorimotor and cognitive performance powers in the outside world, only their sensorimotor *input and output contact points* with our bodies are part of our cognitive (= mental) state, not the parts that extend beyond.

(17) Perhaps it could be otherwise too, as in the case of a hypothetical TT-robot whose generating mechanism is indeed partly located outside its body: Maybe parts of our brain could be removed and still functionally integrated with the rest wirelessly, through telemetry or some other action at a distance: But that would just be a widened, spatially distributed *body*.

(18) The resultant distributed cognitive state would still not be the same thing as considering a telescope, car, library or calculator as parts of a distributed cognitive state (for either a human or a TT robot): Those would still just be parts of the sensorimotor I/O to and from the cognizer's body.

(19) We are not aware of the generating mechanism underlying our cognitive capacity, however, only of its outcome: Hence retrieving a word from memory or retrieving a word via a Google search feels much the same to us.

(20) Does the fact that cognizing is a conscious mental state, yet we are unconscious of its underlying functional mechanism, mean that the underlying functional mechanism could include Google, Wikipedia, software agents and other human cognizers' heads after all? That question is left open for the reader.

(21) The worldwide web, a distributed network of cognizers, digital databases and sofware agents, has become our "Cognitive Commons," in which cognizers and cognitive technology can share cognizing anytime and anywhere, and interact globally with a speed, scope and degree of interactivity that yield distributed cognizing with performance powers inconceivable within the scope of individual cognition.

(22) Such changes go beyond mere quantitative increase in efficiency and performance power. As we increase our use and reliance on cognitive technologies, they effect and modify how we cognize, how we do things and what we do. Just as motor technology extended our physical ability and modified our physical life, cognitive technology extends our cognitive ability and modifies our mental life.

## Part I: What Distributed Cognition Is Not.

**Meaning: Narrow and Wide.** Philosophers, in wrestling with the problem of meaning ("Is meaning in the head or is it in the world?") have sometimes resorted to saying that there are *two* kinds of meaning, "narrow" and "wide," the former located between the ears and the latter distributed across the entire universe -- both the Newtonian universe of distant stars and the Platonic universe of the eternal truths of logic and mathematics. The wide meaning of "apple," for example, includes not only whatever it is that I may have in mind when I think of or say "apple," but also what apples really are, out there in the world.[1]

That, however, is all metaphysics, and concerns the existence and "reality" of some elusive entity called "meaning." The mission of cognitive science is more modest: Humans and other organisms have certain functional capacities, including metabolism, reproduction, and locomotion. It is clear that each of these capacities is "narrow," even though it sometimes involves a local interaction between the organism and part of the world around it (be that other organisms or the inanimate world).

**Performance Capacity.** Movement itself, inasmuch as it includes the movements of parts of the organism, and not just the whole of the organism, covers everything that we are able to *do*; and that, in turn, extends naturally to all of our cognitive capacities – what we are able to think, deduce, understand, etc. – encompassing also the internal mechanisms that generate those capacities.

So far, that makes all of cognition narrow: skin and in. It is not that we do not (as in metabolism and reproduction) interact with objects (and skins) outside our own skin. Although usually it is not particularly illuminating to speak of eating and digestion as a dyadic function, "distributed" between predator and prey. Reproductive function is for the most part decidedly dyadic and to that extent distributed more widely than a single organism.[2]

**Distributed Perception?** Gibson (1966), too, has stressed that even something as seemingly passive as *seeing* is in fact interactive, with the locomotory organism perceiving things in terms of their sensorimotor "affordances" – what our dynamic bodies are able to *do* with external things. Don't look for the purely sensory property that all "chairs" share: their real invariant is that they afford "sittability upon" – a property that cannot even be defined without reference to the shape and motor capacity of our bodies as well as the shape of things in the external world.

Does it follow from this that the perceptual state of perceiving something is a *distributed* state that includes the perceiving organism as well as the external object or event or action that is being perceived? And – to extend this question further – is the cognitive state of thinking or knowing about something a distributed state, consisting of the cognizing organism plus the external object or event or action (or property or state) that is being cognized (Clark & Chalmers 1998; Wilson 2004)?

---

[1] This is even more evident when it comes to what is meant by "superstrings" or "prime number."

[2] In the case of sexual reproduction, *ab ovo*, and in the case of asexual reproduction, *a posteriori*, so to speak. Indeed, there is perhaps a lesson to be learned about cognitive function from the two forms of reproductive function, since both are "productive" of something beyond the narrow borders of the particular organism in question.

Let us defer reply until we consider a few more cases, noting only that this question about whether perception/cognition is just (i) internal and local or (ii) internal/external and distributed is similar to the question of whether meaning is narrow or wide.

**Physical States: Narrow and Wide:** A trivial answer would be that *every* physical state is "distributed" in that nothing is ever causally isolated from everything else. So in singling out ("individuating") any physical "state" we are individuating an arbitrary subset of the total state of the universe: This chair is not causally isolated from the ground it rests upon, nor the ground from the rest of the planet, spinning about the solar system, etc. By that token, all states are wide – as wide as the world, including oneself, sitting on the chair.

But the fact that there is no such thing as an absolutely isolated local entity or state is not what we mean when we ask whether cognitive states are narrow or wide. Otherwise, the state of a toaster, toasting bread, is wide too, and includes not only the toaster and the bread, but also the events transpiring on faraway Alpha Centauri.

But, leaving aside the physics and metaphysics of wide causality and action-at-a-distance: what about just the toaster and the bread? Does the "state" of a toaster, toasting bread, include the bread, being toasted? It seems obvious that this distinction, too, is arbitrary, hence trivial: We can include the toaster in a distributed hybrid state and call that a state of the toaster, or a state of toasting. Or we can say that the toaster does what it does, and the bread gets done to it whatever is done to it, but we will consider their states as distinct, acting upon one another (more the toaster acting on the bread than vice versa, unless the toast catches fire) but not a joint, distributed state worth speaking of as such, in useful discussions of either toasters or bread, and their respective functional states and properties.

**Autonomous Systems.** But although there are no states or systems that are completely isolated causally, there are surely "things" – like chairs and bread and toasters -- that are sufficiently isolated to be called *autonomous* things. Some of these autonomous things will be (again, only relatively) static, like chairs and bread, and some relatively dynamic, like a toaster – if plugged in and functioning. Some of these autonomous things may also be parts of other, wider autonomous things. Toasters have functional parts that can do what they do on their own, in isolation from the toaster. A toaster, in turn, may be part of a more elaborate device that toasts as well as butters, fills and wraps your sandwich; or simply a component in a modular commercial kitchen.

So in this approximate way, bracketing the issue of wide causality, we arrive at the notion of autonomous systems, like toasters, composed sometimes of components that are themselves autonomous systems. Let us call those subcomponents autonomous *modules*, and note that any autonomous system could in principle also be an autonomous module in one (or many) wider autonomous systems.

But is a toaster really autonomous? Don't we have to build it, plug it in to the electrical system, and then put in the bread, and set the level, etc.? Are the toaster and bread and ourselves just part of a still wider distributed system, the one with the real autonomy, while the toaster and the bread are merely "slave" systems, with no autonomy of their own?

We cannot avoid extending our relentless questions to asking what we really mean by "autonomy": Is anything really autonomous, apart from the universe itself, or God almighty? This is again the question of causal isolation, and maybe we can again finesse it by settling for commonsense approximations: A system is autonomous if it can do what it does "on its own." It's just that systems differ in what they can do on their own. A toaster is an autonomous system that can only toast bread -- and that, only if a person plugs it in, puts in the bread, presses the switch. A person is an autonomous system that can (among other things) plug in a toaster, put in bread, and press the switch. And so it goes. Both autonomy and functional capacity look modular, and superordinate autonomous systems may include the distributed modularity of many component autonomous systems.

We can easily get lost in this mereological maze, so let us avoid the lure of "general system theory" and just note that, yes, there are quasi-autonomous things and quasi-autonomous states, and those things and states may themselves be distributed parts of other, wider quasi-autonomous things and states. That's all indisputable even before we get to the question of cognizing and distributed cognition. But before we broach that question, we alas have to ask yet another basic question: What is cognition?

**Cognitive and Vegetative Function.** To a first approximation, we have already answered this: cognition is whatever gives cognitive systems the capacity to *do* what they can do. It is the causal substrate of performance capacity. Cognitive systems ("cognizers") include ourselves and perhaps other animals (and possibly also extraterrestrial creatures, if they exist). Do they include anything more? Are living systems the only cognizers? Are cognizers necessarily local, or can they be distributed? And is the capacity underlying *everything* that we cognizers can do cognitive, or only the capacity underlying *some* of what we can do?

One question at a time. Let us first agree that not everything a human being can do is cognitive. Breathing, for example, except in some special cases, is not cognitive; neither is balance, again, except in some special cases. What are the special cases? They are when we breathe or adjust our balance consciously. Otherwise, breathing and balance are unconscious and automatic – we might call them "vegetative" rather than cognitive functions.

**Consciousness.** But surely consciousness itself cannot be the mark of cognition either, because although when we take conscious control of our breathing or our balance that is undoubtedly cognitive, we are not really conscious of *how* we control breathing or balance. If we suddenly feel we are suffocating or falling over, we "command" our lungs to breathe and our limbs to right themselves, but we are hardly conscious of how our commands are implemented. It is physiologists who must discover how we manage to do those things.

And, by the same token, if we do something that we are more accustomed to calling cognitive, such as perceiving a chair, understanding the meaning of a word, or remembering the product of seven times nine, all of which we cognize consciously, we are nevertheless unconscious of *how* we manage to perceive a chair as a chair,[3] how we understand the meaning of, say "cognitive," or even how we retrieve (from wherever we "stored" it decades ago) the product of seven and nine.

Fear not, dear reader, we have not been forced into the clutches of the metaphysical problem of Free Will here. We simply need to make the observation that what makes some of our capacities cognitive rather than vegetative ones is that we are conscious *while* we are executing them, and it feels like we are causing them to be executed – not necessarily that we are conscious of *how* they get executed (Libet 1985).

But that's not enough. We have an initial approximate criterion for what performance capacities count as cognitive: It is the ones we execute consciously, which just means that we normally have to be conscious while we are executing them (This criterion is actually flawed, but we will fix it later.)

**Is There Cognizing Without Consciousness?** Now the second question: Are there any other cognitive systems besides ourselves and animals? We have already noted that not all of our performance capacities are cognitive: the cognitive ones are the ones we execute consciously (although we are not conscious of how they are executed by our brains). The question of whether systems other than animals like us cognize is hence related to the question of whether or not there can be cognizing without consciousness: It concerns which organisms are conscious, and whether nonconscious -- perhaps even nonliving -- systems, can be cognizers too.

**The "Other-Minds" Problem.** Let us quickly agree (with Hume and Descartes) that there is absolutely no way for one to know *for sure* whether anyone (or anything) but oneself is conscious. (This is called the "other minds" problem, and it is insoluble.) Hence we already have a problem here, if consciousness is the mark of the cognitive. We can't know for sure who or what is or isn't conscious. But do things get even worse? Doomed to be left agnostic about whether anyone or anything else is conscious, are we even more agnostic, then, about whether nonconscious systems can cognize?

---

[3] i.e., how to detect its Gibsonian "affordances."

**What Is Alive?** Here there may be useful lessons to be drawn from the problem of life: Very similar questions have been raised about what it is for a system to be alive. We used to think there had to be a "vital force." Now we know better; life is just the state of certain dynamical systems, having certain structural and functional properties, including molecular ones. The properties of living systems are all objective and observable, so once it has been ascertained that those properties are indeed present, there is no vitalist homologue of the "other minds" problem to trouble us, about whether or not the system is really in a biotic state, i.e., "really alive." The observable, objective properties of living systems exhaust all there is to being alive (other, perhaps, than the "other minds" problem itself, for those who hold that all living systems must be conscious!).

**Biotic States: Narrow and Wide?** The same question of distributedness – "narrow" versus "wide" life – arises also with living things: We all know the case of the amoeba, which is an individual, autonomous, one-celled organism, definitely alive in its own right. But when individual amoebae find themselves together in a certain chemical gradient, they coalesce and become a further, superordinate, fungus-like organism called a slime-mold. This is "distributed" life, in the sense that it is wider than any of the individual amoebae (who nevertheless remain alive too), and encompasses the entire slime mold, which is then an autonomous, superordinate, living organism.

Something like the slime mold was also probably the origin of all multicellular organisms, all the way up to ourselves: We are instances of "wide" life, distributed over all our individual living (though only minimally autonomous) cells.

**Distributed Life.** So far, so good. But, can a group of organisms working and functioning as one, be an individual organism? What about a sports team or an army unit? Or a colony of ants or bees? Even more controversially, some have gone on to argue that an entire biological *species* may also be an individual organism -- a wider, superordinate organism, distributed over all its members, much the way the slime mold is a superordinate organism distributed over its individual modules, the amoebae (Hull 1976).

And it can get even wider, some arguing that Earth itself is a superordinate organism, "Gaia," distributed over the entire biosphere (Lovelock 2000). Perhaps some exobiologists will want to argue that if there is life elsewhere in the universe, then all instances of biotic systems are distributed subcomponents of yet another individual mega-organism.

We will not settle the question of "distributed life" here one way or the other, except to note that (apart from the relatively coherent multi-cellular organisms "supervening" on individual living cells) the criteria for individuating wider and wider forms of life begin to look just as arbitrary as the extension of physical states (on the grounds that no physical sub-state is totally isolated causally) to the size of the entire universe. Nor is it clear any more what, if anything, is at stake when we can call many distributed things one superordinate thing at will, mixing and matching according to taste. We should try to avoid such a state of affairs with distributed cognition.

The questions of distributed life and of distributed cognition, however, are not independent, because (to a first approximation) it is living organisms that cognize (those of them that do), and it is likewise living organisms that are conscious (those of them that are).

**Spatial and Causal Disjointness.** Up to the level of continuous multi-cellular organisms, we can agree about what is and is not a living organism. Consensus and coherence collapse only when we move to the level of the species or Gaia, as both the spatial distance and the causal interactions among the component organisms get distributed more and more widely and loosely: Fungi are the biggest uncontestable organisms on the planet. Some of them can grow underground to a size of over 2,000 acres and live more than 2,000 years. Their individual fruiting bodies, the mushrooms (that we mistake as being the whole organism), though remote from one another in space, are all continuously connected.

What about a coral colony, or, better, an ant colony? Is it such a stretch from the spatially continuous and tightly coupled causal interactions of the amoebae that constitute a slime mold to the only somewhat more spatially disjoint and less tightly coupled causal interactions of the ants that constitute a colony? Within multi-cellular organisms there is action at a distance (for example, via chemical gradients) as well as coherent but distributed activity (as in a neural network). And we all know that "spatial continuity" breaks down at microscopic scales.

Fortunately, in individuating organisms there are other criteria besides spatial and temporal continuity. There is DNA, which can help resolve (up to cloning) whether or not two bits are (or were) indeed parts of the same organism. But genetic relatedness is only relative, which is what allows some to argue that species are individuals and that Gaia is a mega-organism.

**Distributed Mental States?** We can avoid having to wrestle with the metaphysical problem of individual identity in making our bets as to whether something is a case of individual life or just multiple life, interacting. Siamese twins offer a clue: Why are we ready to contemplate the possibility that Gaia, or an entire species, or an ant colony, might be one single, widely distributed, physically disjoint organism, yet we are not ready to consider that Siamese twins, no matter how tightly fused they are physically, are one single organism? The example illustrates how tightly interconnected and fused the questions of distributed life and distributed cognition really are (at least in our minds):

The reason we would never dream of saying that Siamese twins are one single distributed organism is that *they have two different minds*. And distinctness (or identity) of minds trumps all of our other intuitions and inclinations, insofar as individuating either organisms or cognizers is concerned. Suppose Siamese twins could share every last body part yet could still have two distinct minds: not as in the ambiguous case of multiple personality disorder, where the "minds" (if they are really different minds at all) come and go serially, like masks, but where they are always jointly present, and you can communicate with them, and they with one another, simultaneously, exactly as in the case of ordinary Siamese twins (except that both cannot speak at exactly the same time). We would still have profound difficulty seeing them as one and the same "organism" – or perhaps we should say that the notion of an individual "organism" would simply lose its meaning for us in such a situation: They would be one "biotic system," in some technical sense, but two distinct "cognizers."[4]

Is this "animism" (which was probably always latent in the "vitalism" that has since been discredited by molecular and evolutionary biology), just in our minds? Should we be accepting objective, system-based functional inventories of what does or does not count as a distinct cognizer, as we do with what does or does not count as being alive? Or does our subjective sense have some privileged say in the matter?

**Mind-Reading.** The very same mentalistic intuition that underlies how many cognizers we perceive within a single organism can cut the other way too: The reason most of us are not ready to see an ant colony, a species, a corporation, a government, a sports team, an army unit or Gaia as either an individual organism or a cognizer is that we do not perceive any of them as having a mind. We can, with a little effort, see a tree or a fungus, a coral, an amoeba or a slime mold as a living organism, like us, especially if biologists tell us it is so; but we would have great difficulty seeing any of them as cognizers – unless we are ready to see them as being conscious (having a mind).

We do have natural "mind-reading" abilities and inclinations (Whiten 1991), as along with fertile imaginations. When we are children, and our animism is at its "widest," we are ready to see a tree as watching or even waving at us, or to believe that it hurts the tree when we kick it. Past a certain age, children also have a great deal of difficulty *not* believing that it hurts a dog, or another child, if they kick it.[5]

We perceive other minds because we can (sometimes) detect the Gibsonian "affordances" (perhaps via our "mirror neurons") of being in a mental state: We know what it's like to have a mind, because we each have one. The rest is our "mirror neurons," detecting when another mind is in a mental state like our own, because it is doing something like what we would be doing in that mental state. In other words, we mind-read through a combination of having a mind and perceiving its bodily performance correlates in others (Gallese & Goldman 1998).

[4] In contrast a baby born with extra limbs will always be considered as a single organism, regardless of how many extra limbs it has, as long as it has just one cognizing mind.

[5] Although without moral training, that is not necessarily enough to prevent the child from kicking it!

The "other minds" problem does not go away; our 'mind reading' is not based on flawless deductive reasoning. The logic that similar consequences must have similar causes (I have a mind and know its bodily outcome, therefore such bodily outcomes in others must be caused by a mind) gives raise to false positives and false negatives. Hence we are fallible mind-readers. Seeing the tree as having a mind is probably a false positive. Might seeing Gaia as *not* having a mind be a false negative? Perhaps. But let us be clear about exactly what we would be getting right or wrong, when we made a correct "hit" versus a false positive or negative:

**Living and Cognizing.** If Gaia, or a species (e.g., the earth's elephant population as a whole) did have a mind, that would mean, roughly speaking, that it was the kind of thing that was capable of having a headache (not necessarily having a head, just having a headache), say, a migraine. The migraine is just a stand-in, here, for our intuitions about what it is to have a mind at all. To have a mind is to be in a mental state, and a mental state is simply a *felt* state:  To have mind is to *feel* something – to feel anything at all (e.g., a migraine).[6]

And make no mistake about it: you *must* have a mind – i.e. you must be in a mental state, you must be able to feel -- in order to have any inkling at all of what it means to have a mind! A toaster will not get that from a dictionary definition. Outside minds there is nothing but mindless (feelingless, insentient) functionality.

**The Migraine Test.** The migraine is merely our stand-in for the capacity to *feel anything at all* -- in other words, for being conscious. We all know what it feels like to have a headache. All feelings are pretty much like that, *mutatis mutandis*, from what it feels like to perceive a chair, to what it feels like to understand the meaning of a word or to remember the product of seven and nine. Note that what is essential for having a mind is not having the performance capacity itself – being able to detect the presence of the chair, being able to define or reply correctly to the word, being able to retrieve "sixty-three " -- nor is it essential to have an understanding of the underlying causal mechanism of that performance capacity (knowing how we manage to do it). The essential thing for having a mind is being able to feel what it is like to have and execute the capacity – or to feel anything at all (e.g., a migraine). This is the consciousness that accompanies cognizing (though without necessarily any consciousness of how the cognizing actually works).

Suppose it was somehow *true* that Gaia (or the entire elephant species, or an ant colony) was indeed a superordinate living organism, distributed across everything in the earth's biosphere (or across all elephants, or all the ants in a colony). And suppose the reason we wrongly thought Gaia was *not* an organism was that we couldn't imagine such a distributed system as being capable of having a migraine (or any other mental state). We could of course have been wrong about that too: Maybe

---

[6] Having a mind, being in a mental state, being conscious, being in a conscious state, feeling, being in a feeling state, feeling anything at all -- all of these are synonymous.

Gaia *could* have a migraine. (Because of the other-minds problem, there is no way to be sure one way or the other.)

*But suppose we were right that Gaia has no mind yet wrong that Gaia is not a living organism.* In that case, our mind-reading mirror-neurons would have been right – they detected no mind. But they would nevertheless have steered us into a false negative, because Gaia, though mindless, is nevertheless alive. In contrast, the child's mirror-neurons commit a false positive on the migraine test, wrongly inferring that a tree *does* have a mental state, though it does not, but the child is nevertheless right (though for the wrong reasons) that the tree is alive. In both cases, being an organism was conflated, animistically, with having a mind. This is an error; living and feeling are not necessarily the same thing. There can be living organisms that have no mental states and there can be nonliving systems that do have mental states.

**Cognitive States and Mental States.** Can the same distinction be made, however, if we apply the same mind-reading criterion to being a *cognizer*, rather than to being a living organism? We (or rather, our mind-reading mirror neurons) insisted, in the case of the Siamese twins with only one body, that even if Biology were to tell us that they were one single organism, they would still be two distinct cognizers, if they had two distinct minds: They would not have one, shared mind, even though they did have one, shared body. And if they had a migraine, it would be *two* migraines, even if it was implemented in one and the same head -- just as when something is a "headache" for the US Congress, it is at most N distinct headaches in the heads of N distinct members of congress, with no further superordinate entity feeling an N+1$^{st}$ headache (or feeling anything at all). There is no such thing as a distributed migraine – or, rather, a migraine cannot be distributed more widely than one head. And as migraines go, so goes cognizing too -- and with it cognition: Cognition cannot be distributed more widely than a head -- not if a cognitive state is a *mental* state.

**Cognitive Technology: Tools R Us?** Does this settle the question of distributed cognition, or does it beg it? The case for distributed cognition is based mostly on *cognitive technology*: the argument is that even something as simple as an external piece of paper with a phone number on it is a piece of cognitive technology -- a peripheral device on which data are stored. If the phone number were encoded inside one's brain, as a memory, there would be no dispute at all about its being part of the (internally) distributed cognitive state of, say, knowing or finding that phone number. Why, then, would we no longer consider that same datum as part of that distributed cognitive state just because its locus happened to be outside the cognizer's body?

Moreover, once we realize that our cognitive states include data that are located on an external piece of paper, then it becomes apparent that they include far more than that -- widely distributed things, ranging from everything located in our libraries and on the Web, to every auxiliary device, process or datum that may enter into any cognizing or its outcome, extending also to everything located in the narrow heads of all other individual cognizers (Hollan et al. 2000). Whether we want to include in a cognitive state everything that can potentially enter into anyone's cognizing or only what actually enters into someone's cognizing, either way, on this extended view, cognition is looking exceedingly wide.

**The Extended Mind.** This wide view of distributed cognition is also called the "extended mind hypothesis" (Clark & Chalmers 1998; Wilson 2004). It has some affinities with wide theories of meaning, in which apples themselves, or the truths about apples, are taken to be part of the distributed meaning of "apple," which is thereby extended beyond what may be going on within the narrow confines of the head of any individual, local cognizer. The extended mind is also reminiscent of the notion of "wide" toaster-states (in which the bread is part of a distributed state of the toaster), thereby also inheriting the apparent arbitrariness of such state extensions.

We must accordingly ask ourselves why we would want to contemplate such arbitrary extensions of what it is to have or to be a mind, hence to be a cognizer and to cognize? Why would it even cross our minds? The answer is again the (insoluble) other-minds problem: Since there is no way of knowing for sure whether any cognizer other than oneself has a mind, there is even less way of knowing whether or not there can be cognizing without a mind, or even of knowing what the actual geographic boundaries of a mind are.

We know, objectively, what cognition *does*. Doing is performance capacity. Cognitive science is also beginning to make some progress in explaining the functional mechanism generating that performance capacity (i.e., how our brains do it). We also know that so far the only systems that our adult "mind-reading" capacities have tentatively identified as being cognizers have been living organisms like ourselves. Our confidence that we have detected an "other mind" increases the more the candidate resembles ourselves, both in its appearance and in its performance capacities. That is in fact the (narrow) meaning of "cognition": the kinds of things that I and other living organisms can do, using our minds.

But there are other candidates that also seem to be able to do some of the things that living organisms like us can do, and not just the vegetative things, but the things we have identified as cognitive, when performed by us. Computers and robots are not only doing more and more of what only living organisms like us used to be able to do, but even the functional mechanisms that biology and cognitive science are proposing to explain how organism*s* do it often turn out to draw on the same functional mechanisms that explain how computers and robots do it. (Indeed, the functional explanation often comes from the fact that we have programmed computers and designed robots to do what we do, and in so doing, we have also provided a potential explanation of how our own brains do it). So, if it walks and quacks like a duck, and even its internal mechanism is like that of a duck, it's only natural to assume it's some kind of duck too.

**The Turing Test.** This is the rationale and the methodology behind the Turing Test (TT; Turing 1950; Epstein et al. 2008): If we can design a system that can do everything that we can do -- and do it so well that even our mind-reading mirror-neurons perceive it as having a mind -- then we have no more (or less) reason to doubt that it has a mind that we have for doubting that other human beings have minds (again because of the other-minds problem). In addition, the TT-passing candidate – which has to be a robot, because a computer alone cannot have all of our sensorimotor capacities – will provide us with at least one explanation of the functional mechanism underlying our own cognitive capacity.

Hence the question of narrow vs. wide cognition is also a question about what is and is not part of the functional mechanism of a TT-passing robot: What needs to be "inside" such a robot in order to pass the TT? Note that this is not the question of whether we need to pack all its functions inside its head, the way ours are packed inside our heads. It is conceivable that the mechanism of the TT robot could be more widely distributed: some of it inside and some of it outside its body, integrated wirelessly, perhaps, from some central location. The states consisting of the joint activity of the robot-internal and the robot-external components of the mechanism that give the robot the capacity to pass the TT would be indisputably distributed cognitive states.

But those hypothetically distributed robot states (if they are possible at all) do not settle the question we are inquiring about here. Nor would they settle it even if it were somehow possible to breed people with parts of their brains physically located outside their bodies and their joint activity integrated through wireless telemetry or some such. Such a hypothetical distributed robot (or person) could even have a distributed migraine. But what we would really have then would be a robot (or person) with an extended (or distributed) *body*. The constituents of its mental states would all still be (distributed) within that one distributed body. Our brains, after all, are still parts of our bodies, even if they could be removed, all or in part, temporarily or permanently, the way our hearts have been, and somehow kept functionally integrated with our bodies wirelessly.

This is all cog-sci-fi. But the point of the example is to show that this sort of hypothetical cognitive state – distributed across multiple parts of a robot's functional mechanism (or even multiple parts of an organism's brain) that happen to be widely separated in space but coordinated wirelessly (Dennett 1981) – does not address the question of whether or not *cognitive technology* is part of our cognitive state too. An affirmative answer to the question of whether, if the parts of my brain that control the left and right sides of my body could be moved out of my brain and two miles apart, while still being able to remotely coordinate my walking, does not address the question of whether cars or calculators are a part of my mind. [7]

---

[7] Please note that if cognitive technology is only a tool and not part of our mind and a cognizer itself, this does not imply that it does not have profound effects on how we cognize --more on this later.

**Software Agents.** It is not just today's (sub-TT) robots that appear to be doing some of the things we cognizers do: Software agents seem to be doing it too, including communities of distributed software agents, interacting among themselves, trawling the net, executing local as well as distributed computations on local and distributed data, and displaying -- not just individually but also collectively -- performance capacities that, in living organisms like ourselves, we would have taken to be the result of cognizing (Dror 2007).

These autonomous devices – both hardware and software -- are, of course, like toasters. But they are "cognitive" toasters, in that they operate not on bread but on informational inputs, to generate, as output, performance that we *would have called cognitive if we had been the ones doing it*. Particularly in the case of the joint activity of distributed software agents, autonomously crawling the web, it is obvious why the question of whether a datum is internal or external becomes arbitrary. The datum may be the input to one agent or the output of another, and the distributed cognitive system consists of all the agents and their inputs and outputs together anyway. It makes little sense, nor is it of much use, to try to say which is the bread and which is the toaster in such cases. But is what these systems are doing (whether they are local pieces of hardware or distributed digital data and the software agents programmed to process them) *cognizing*, or just something that is similar to what ordinarily requires cognizing to do? The question seems to be as undecidable as whether or not Gaia is really a living organism.

## Part II: What Distributed Cognition Is.

**Wide-Body Beings.** In Part I we argued that inasmuch as cognition is mentation (i.e., insofar as cognizing is thinking), there can only be distributed cognitive states where there can be distributed mental states. Within the head there are narrowly distributed cognitive states, since neural states are presumably not all local and punctate. If the mechanism that generates mental states and bodily performance capacity (normally the brain) could be more widely distributed in space (beyond the head), and still be integrated somehow so as to generate coordinated mental states and bodily function, then that too would be widely distributed cognition, whether in a hypothetical TT-scale robot or a hypothetically re-engineered organism, but that would also be a *widely distributed body*. Distributed cognition would still not be wider than the body.

Can there be distributed cognition beyond the bounds of the body and the brain? In particular, can external cognitive technology serve as a functional part of our cognitive states, rather than just serving as input to and output from them?

**Mental States Are Conscious States.** Let us consider brain states, rather than just mental or cognitive states. We have agreed that not everything our bodies do is cognitive. Some of it, like breathing, balance, or temperature control is vegetative. So, too, are the brain states that implement those vegetative functions. We have also agreed that although cognizing is conscious, we are not conscious of how cognizing is implemented. When we recognize a chair, or understand a word, or retrieve the product of seven and nine from our memory, the outcome, a conscious experience, is delivered to us on a platter. We are not conscious of *how* we recognized the chair, understood the word, or retrieved “63”. Hence the brain states that implement those cognitive functions are not conscious either. Are *unconscious* brain states mental?

**Are Unconscious Brain States Mental?** The natural answer would seem to be: no. Unconscious states are unconscious states. The states of a toaster are unconscious and certainly not mental. Until further notice, “conscious states” is synonymous with “mental states.” The brain states implementing vegetative function are not mental either: Presumably a person in a chronic vegetative state is as unconscious as a toaster (although, because of the other-minds problem, we can never be sure about either the toaster or the comatose person). The only reason we want to call the brain states that occur while we are conscious *mental* states is that they occur while we are in a conscious state, and they physically implement that conscious state. (We are on the fuzzy boundary of the mind/body problem here.) But just as vegetative states such as the regulation of breathing, which occur unconsciously while we are conscious, are nevertheless not themselves mental, nor part of our mental state, why would we want to call the unconscious state that “delivers” our conscious mental state mental?

When you say to yourself “what is seven times nine?” and then “sixty-three” pops up, you are certainly conscious of thinking “sixty-three.” So that’s definitely mental; and so is the brain state that corresponds to your thinking “sixty-three.” But what about the brain state that actually *found and delivered* the “sixty-three”? You are certainly not conscious of that, although you were just as conscious *while* your brain was finding and delivering “sixty-three” as while you were breathing, though you don’t feel either of those states.

**Neural vs Google Storage and Retrieval.** Let us make the retrieval interval longer then, just to make the problem more vivid: You are trying to remember the name of a poet. You know he wrote “Tell me not in mournful numbers, life is but an endless dream” and his name is on the tip of your tongue, but you just can’t retrieve it. You go to sleep, and next morning “Henry Wadsworth Longfellow” immediately pops up. You were not even awake during the brain state that retrieved it. So what difference does it make if you recall it through an unconscious retrieval state in your brain, or by Googling it (again relying on a state in some remote computer and database of which you are not conscious)? Are they not both based on an unconscious, nonmental state, in the first case narrow and neural, inside your brain, in the second case wide and computational, distributed between your brain and a computer hundreds or thousands of miles away?

**Distributed Databases.** And what about a modern child, who has never bothered to memorize the multiplication tables, as you did, because a computer is always at hand? The only way he ever retrieves 7 x 9 is to key it in, and read off the product. He blindly consults his computer when you blindly consult your memory: What's the difference? Never mind computers: the poet's name could be read out of a static book that indexes poems' first lines. Or you could just ask somebody who knows to tell you who wrote those lines. What difference does it make if the database in which the datum is stored, outside your awareness, is in your brain, or on the shelf of a library, or in someone else's brain?

**Offloading Brainwork.** The beginning of cognitive technology was surely language, which allowed cognizers to "offload" a lot of brainwork onto other brains that could do it for you, and deliver you the results, and vice-versa (Cangelosi & Harnad 2001; Dascal 2004). Are our own neural states, plus Google states, plus book states, plus the neural states in the heads of other cognizers all parts of distributed cognitive states – and if so, *whose* cognitive states? I am presumably the cognizer of my narrow cognitive states, but who is the cognizer of the wide ones?

Or are cognitive states just sui generis, rather than belonging to anyone in particular? Neural firings in brains, plus keystrokes on computers, bits coursing across fibre optic cables, remote disk activity, print in a library book, neural states in other people's brains – all just parts of wide, distributed, disembodied cognitive states, taking place here, there, and everywhere: cognizing, with no cognizer?[8]

At the very least, we need to pinpoint the cognizer of the distributed cognitive state. Let us say it is the *user* of the cognitive technology, and that what we are asking is whether the technology outside the body is part of or merely I/O to/from a narrow cognitive state inside his brain?

**Sensorimotor Technology and Augmented Reality.** Let us start by considering a kindred kind of technology, perhaps not quite cognitive, only *sensorimotor*, with the corresponding states being sensorimotor states rather than fully cognitive ones: Sensorimotor technology probably began in our species' prehistory with tools and weapons, which extended our performance capacities dramatically. Let us consider a relatively recent tool:

If you look at a star through a telescope, is that a distributed sensorimotor state, consisting of your brain and retina plus the telescope (and perhaps also the star), in which your visual capacity is augmented by the telescope's power of refraction? Or is it just *input* to your narrow, skin-and-in sensorimotor state – input augmented by the telescope?

---

[8] (rather like a distributed life, with no organism living it; or a distributed migraine, with no one experiencing it)? Isn't cognition with no cognizer cognizing it like a feeling with no feeler feeling it?)

If you are driving a car, is that an extended sensorimotor state, in which your body is moving at speeds in excess of what it can manage alone, narrowly? The wider, distributed sensorimotor state might include the car and its locomotor capacity. Or is it just *output* from your narrow, skin-and-in sensorimotor state (in this case a slow movement of your foot on the pedal) – output augmented by the horse-power of our external vehicle?

Another example would be operating a crane, and the extended power to reach and manipulate objects that are too far, big and heavy to be manipulated in your narrower sensorimotor state. Is this a widely augmented I, or just I/O to narrow old me?

But before we dismiss too quickly the notion of a wider sensorimotor state, note that some of us have literally experienced a change in our felt body image when driving a large car: Our sense of our own width, pulling through a narrow squeeze, extends to the width of our car, not just our narrow body. This change in body image is not unlike the effect induced by distorting prisms, Virtual Reality, or even surgery, prosthetic limbs, and neurological re-adaptation. Tadpoles mutating into frogs and caterpillars mutating into butterflies might be undergoing similar sensorimotor changes in their body images and powers because of real changes in their (narrow) bodies: Are technology-extended bodies all that different?

**The Advent of Language.** The effects of cognitive technology can be similar to those of sensorimotor technology. Language evolved neurologically for speech and its interactive tempo. We can accelerate recorded speech technologically beyond the rate we can speak it, yet still understand it. Beyond a certain speed, speech becomes gibberish – yet we can read and understand written language at far faster speeds (probably because hearing is a more serial medium of processing and vision is more parallel).

It is virtually certain that there was no specific neural adaptation for reading, which was a technological invention of less than 10,000 years ago. In contrast, the language areas of our brain were shaped genetically several hundred thousand years ago, altering our neural hardware and radically extending our cognitive powers. If spoken language widened our cognitive powers biologically, didn't reading and writing widen them technologically in much the same way?

**Language As Distributed Cognition?** Is language itself a form of distributed cognition? How does the knowledge in other people's heads, conveyed to us auditorily, differ from the knowledge in books, conveyed to us visually? Both allow us to access information without needing to gather it the hard way, through our own direct, time-consuming, risky and uncertain sensorimotor experience. Writing and speaking also allow us to offload our knowledge and memory outside our own narrow bodies, rather than having to store it all internally. Individual cognizers write books, but Wikipedia, for example, seems to be growing spontaneously according to an independent, collective agenda of its own, more like the joint activity of a colony of ants.

Computers, distributed digital databases and automated algorithms have augmented both the speed and the computing power of our brains, and that newfound speed and power is capable of inducing changes in our mental self-image not unlike the ones that sensorimotor technology can induce in our body image: If being deprived of one's spectacles or one's automobile feels rather like the loss of eyes or limbs, being deprived of one's computer or cell-phone feels like the loss of one's intrinsic cognitive and communicative capacity.

**Interactive Cognition.** Human discourse is certainly *interactive cognition*, indeed *collaborative cognition*, and the speed and distance at which we could speak, and understand our interlocutors, set biological limits on the rate and scope of that collaborative cognition, hundreds of thousands of years ago. The speed of verbal thought probably co-evolved with language and is probably of the same order of magnitude as the speed of oral speech. Although reading speed is much faster than listening speed, writing (and typing) speed is not as fast as speaking (and the turn-around time of letter delivery is certainly slower than conversational speed). So in *real-time* interactions, at the speed of thought, we still prefer to talk rather than write.

It is only recently that cognitive technology (in this case, email and texting) has accelerated the potential speed of written interactions in almost real-time to something closer to the speed of thought. Web-based threaded discussion lists, and especially their quote/commentary capability, not only accelerate this interaction still further – allowing individual cognizers to *interact with the text itself* in real time. They also increase the scope of this almost-real-time interaction among distributed minds and distributed texts; and global posting and immediate accessibility potentially make the collaboration almost instantaneous (Harnad 2004).

If the human brain was biologically optimized for interactive cognition at speaking speed, and writing technology slowed down that interactive cycle (in exchange for the other benefits of the transmission and archiving of a written record) for thousands of years, then digital online technology has now once again accelerated the interaction to the speed of thought, increasing its power and productivity by orders of magnitude, and distributing it globally and instantaneously. It is this newfound interactivity (not passive radio, television or film) that is at last truly turning Gaia into McLuhan's (1962) "global village."

**Cognizers and Tools.** So where does this leave the question of distributed cognition? It is still cognizers who cognize -- the tool-users, not the tools.[9] Yet there

---

[9] Did some of the ambiguity arise from the fact that we fell into the habit (perhaps because of funding agency contingencies) of overusing (for funding purposes) a rather vague and equivocal noun and adjective – "cognition" and "cognitive" – instead of a less impressive verb and gerund – "cognize" and "cognizing" – to ask these questions that are basically about thinking and knowing? Would we have

is no doubt that cognitive technology has radically widened the scope of human cognizing[10]. Could "cognitive technology" be the brain's way of off-loading some of its otherwise far greater encoding and processing burden? If so, then the worldwide web, a distributed network of cognizers, digital databases and software agents, has effectively become our "Cognitive Commons," in which distributed cognizers and cognitive technology can interoperate globally with a speed, scope and degree of interactivity that generate cognitive performance powers that would be inconceivable within the scope of individual local cognition alone.

**Cognitive Technology and the Human Mind.** Is cognitive technology limited to increasing the cognitive performance capacity of its users? No. We have argued that cognitive tools are not themselves cognizers, nor do they have -- or serve as distributed substrates of -- mental states. But their effects go well beyond making human cognition more efficient and productive. Just as noncognitive technology (cars, planes, machinery) transformed our somatic lives, so the offloading of brain function onto cognitive technology is now transforming our cerebral lives. Physical technology altered the frequency, intensity, and manner of our muscle use, altering our muscular development (even introducing new 'technological diseases', such as carpal tunnel syndrome). Cognitive technology will do likewise, but instead of affecting our muscles it will affect our brain development, organization and capacities. Changing how we think, learn, and communicate, our cognitive tools are reshaping our minds.

---

been ready to say that a library was doing "distributed knowing," or that it was part of a "thinking state" distributed across brains and book-shelves? Or that "collaborative cognizing" was any more "distributed" than collaborative thinking or knowing (or worrying)?

[10] Epigenetics is perhaps a biological precedent for this (Waddington 1942): Not every trait of an organism needs to be genetically encoded in its ("narrow") DNA. If there are stable environmental influences that can be relied upon to "canalize" the expression of genes without having to be written into the blueprint, that takes a needless load off the narrow code, and even allows it to be more flexible toward wider environmental contingencies. (Perhaps the neural counterpart of Eprigenetics should be called "Epinoetics.")

**References**

Cangelosi, A. and Harnad, S. (2001) The Adaptive Advantage of Symbolic Theft Over Sensorimotor Toil: Grounding Language in Perceptual Categories. *Evolution of Communication* 4: 117-142.

Clark, A. & Chalmers, DJ. (1998) The Extended Mind. *Analysis* 58(1): 7-19

Dascal, M. 2004. Language as a cognitive technology. In B. Gorayska & J.L. Mey (eds.), *Cognition and Technology: Co-existence, Convergence, and Evolution* (pp. 37-62). Amsterdam: John Benjamins.

Dascal, M. & Dror, I. E. (2005). The impact of cognitive technologies: Towards a pragmatic approach. *Pragmatics & Cognition, 13 (3),* 451-457.

Dennett, D. (1981) Where am I? In D. Dennett, *Brainstorms: Philosophical Essays on Mind and Psychology* (pp. 310-323). Cambridge, MA: MIT Press.

Dror, I. E. (2007). Land mines and gold mines in cognitive technologies. In I. E. Dror (ed.), *Cognitive Technologies and the Pragmatics of Cognition* (pp1-7). Amsterdam: John Benjamins.

Epstein, R, Roberts, G & Beber, G. (2008) (Eds) *Parsing the Turing Test: Philosophical and Methodological Issues in the Quest for the Thinking Computer* Springer

Gallese, V & Goldman, A. (1998) Mirror neurons and the simulation theory of mind-reading. *Trends in Cognitive Sciences* 2(12): 493-501

Gibson, JJ. (1966) *The Senses Considered as Perceptual Systems*. Greenwood Publishing Group

Harnad, S. (2004) Back to the Oral Tradition Through Skywriting at the Speed of Thought. In: Salaun, J-M & Vendendorpe, C. (Eds.). Le défi de la publication sur le web: hyperlectures, cybertextes et méta-éditions. Presses de l'enssib.

Harnad, S. and Dror, I. (2006) Distributed Cognition: Cognizing, Autonomy and the Turing Test. *Pragmatics & Cognition, 14 (2),* 209-123.

Hollan, J, Hutchins, E & Kirsh D (2000) Distributed cognition: toward a new foundation for human-computer interaction research. *ACM Transactions on Computer-Human Interaction* 7(2): 174-1962

Hull, DL (1976) Are Species Really Individuals? *Systematic Zoology* 25 (2): 174-191

Libet, B. 1985. Unconscious cerebral initiative and the role of conscious will in voluntary action. *Behavioral and Brain Sciences* 8: 529-566.

Lovelock, James (2000). *Gaia: A New Look at Life on Earth*. Oxford University Press

McLuhan, M. (1962) The Gutenberg Galaxy: The Making of Typographic Man. University of Toronto Press

Turing, A.M. (1950) Computing Machinery and Intelligence. Mind 49 433-460

Waddington, CH. (1942), The epigenotype. Endeavour 1, 18–20.

Whiten, A. (Ed.) (1991). *Natural theories of mind: Evolution, development, and simulation of everyday mindreading*. Oxford: Blackwell

Wilson, R. A. (2004) *Boundaries of the Mind: The Individual in the Fragile Sciences - Cognition*. Cambridge University Press